\documentstyle[preprint,tighten,aps,epsf]{revtex}
\begin{document}
\draft
%
%
%
%
\title{
Fermi's golden rule in a mesoscopic metal ring}
\author{
Peter Kopietz and Axel V\"{o}lker}
\address{
Institut f\"{u}r Theoretische Physik der Universit\"{a}t G\"{o}ttingen,
Bunsenstrasse 9, D-37073 G\"{o}ttingen, Germany}
\date{June 1, 1999}
\maketitle
\begin{abstract}

We examine the time-dependent
non-equilibrium current in a
mesoscopic metal ring threaded by a static
magnetic flux $\phi$ that is generated by 
a time-dependent electric field
oscillating with frequency $\omega$.
We show that in quadratic order in 
the field
there are three fundamentally 
different contributions to the current.
(a) A time-independent contribution  which
can be obtained from a thermodynamic
derivative.
(b) A term increasing linearly in time 
that can be understood in terms of Fermi's golden rule.
The derivation of this term requires a careful treatment
of the infinitesimal imaginary parts 
that are added to the real frequency $\omega$ when 
the electric field is adiabatically switched on.
(c) Finally, there is also
a time-dependent current oscillating with
frequency $2 \omega$. 
We suggest an experiment to test
our results.  
\\
{ }
\\
\noindent
Keywords: persistent currents, non-linear response

\end{abstract}
\pacs{PACS numbers: 73.50.Bk, 72.10.Bg, 72.15.Rn}
\narrowtext

\section{Introduction}
\label{sec:intro}

Consider a mesoscopic metal ring threaded by 
a time-dependent magnetic flux $\phi ( t )$ 
that has a static component $\phi$ and
a part that oscillates with frequency $\omega$,
 \begin{equation}
 \phi ( t )  =
\phi + \phi_\omega  \sin ( \omega t )
\; .
 \label{eq:flux}
 \end{equation}
By Faraday's law of induction, 
the oscillating part generates
a time-dependent electric field directed along
of circumference of the ring,
$E(t)=E_\omega \cos(\omega t)$, with amplitude
 \begin{equation}
 e L E_\omega  = 2 \pi \omega \frac{\phi_\omega  }{ \phi_0 }
 \;  .
 \label{eq:Faraday}
 \end{equation}
Here $L$ is the circumference of the ring, $-e$ is the charge
of the electron, and $\phi_0 $ is the flux quantum.
We would like to know the induced current
around the ring. 
In the limit $\omega \rightarrow 0$ this is just the
usual persistent current\cite{Hund38,Imry97}.
But what happens for frequencies in the range
between $10^{8}$ and $10^{13}$ Hz, which for
experimentally relevant rings\cite{Levy90,Mohanty96} corresponds
to $\Delta \ll \omega \ll \tau^{-1}$?
Here $\Delta$ is the average level spacing at the Fermi energy,
and $\tau$ is the elastic lifetime.
We use units where $\hbar$ is set equal to unity.
This problem has first been studied by
Kravtsov and Yudson\cite{Kravtsov93} (KY), who found 
that in quadratic order the time-dependent field induces 
(among other terms that oscillate) a
time-independent non-equilibrium current
$I^{(2)}_0$.
Calculating the disorder average of this current 
perturbatively, KY found that it has the peculiar
property that for frequencies exceeding the
Thouless energy $E_c = \hbar {\cal{D}} / L^2$
(where ${\cal{D}}$ is the diffusion coefficient)
the average of $I^{(2)}_0$ does not vanish
exponentially, but 
only as $\omega^{-2}$.
This is in disagreement with the intuitive expectation that
the external frequency $\omega$ leads to a similar
exponential suppression of this mesoscopic non-equilibrium
current
as a dephasing rate in the case 
of the equilibrium persistent current\cite{Schmid91,Altshuler91}.
The perturbative calculation of KY is based on the assumption of a
continuous energy-spectrum, which means that the level-broadening due to
dephasing, $1/\tau_\varphi$, must exceed the average level spacing at
the Fermi energy, $\Delta$. If we assume that for low temperature $T$ the
dominant dephasing effect comes from electron-electron interactions, a
simple estimate\cite{Voelker99} shows that 
$1/\tau_\varphi ( \omega ) < \Delta$ for
$| \omega | \leq E_c$ in the limit $T\rightarrow 0$. 
Hence, for
frequencies smaller than the Thouless
energy the spectrum is discrete and the perturbative analysis
breaks down. 
In this work we shall
show that in this case 
the term considered by KY
is not constant, but grows linearly 
in time, a result which can be understood  
simply
in terms of Fermi's golden rule of time-dependent
perturbation theory.

It is important to point out the difference between
the current considered here and the
direct current due to the
usual photovoltaic effect.
It is well known\cite{Belinicher80} that
irradiation of a medium without an inversion
center by an alternating electric field can give rise to
a direct current (photovoltaic effect).
The lack of inversion symmetry
can be due to impurities and defects in a finite sample.
For mesoscopic junctions the photovoltaic direct current
has been studied in Ref.\cite{Falko89}.
In this case the average current vanishes, because disorder averaging
restores the inversion symmetry. 
In our case, however, we calculate the direct current
induced in a mesoscopic ring threaded by a magnetic flux. 
Because the magnetic flux breaks the time-reversal symmetry, 
the direct current considered here has a finite disorder average.
Thus, the physical origin of a mesoscopic non-equilibrium current
discussed in this work is quite different from Ref.\cite{Falko89}.

\section{
The 
quadratic response function: What is wrong with the Green's function approach? 
}

We consider non-interacting disordered electrons
of mass $m$ on a mesoscopic metal ring
threaded by the time-dependent
magnetic flux  given in Eq.(\ref{eq:flux}).
Suppose that we have
diagonalized the Hamiltonian 
in the absence of the oscillating
flux (i.e. for $\phi_{\omega} = 0$ in Eq.(\ref{eq:flux}))
for the given realization of the disorder.
The time-independent part of the Hamiltonian is
then $\hat{H}_0 = \sum_\alpha \varepsilon_\alpha 
c^\dagger_\alpha c_\alpha$,
where $\varepsilon_{\alpha}$ are the exact 
electronic eigen-energies  
for fixed disorder, which are labeled by
appropriate quantum numbers $\alpha$.
The operators $c^{\dagger}_{\alpha}$ create 
electrons in the corresponding eigenstates
$| \alpha \rangle$.
If we now switch on the time-dependent part of the field,
the Hamiltonian becomes
$\hat H =
\hat H_0 + \hat V(t)$, with
\begin{eqnarray}
  \label{eq:vt}
  \hat V(t)&=&\frac{2\pi}{m
    L}\delta\varphi(t)\sum_{\alpha,\beta}\langle\alpha|\hat P_x
  |\beta\rangle  c^\dagger_\alpha  c_\beta \nonumber \\
&+&\frac{1}{2m}\left(\frac{2\pi}{L}\delta\varphi(t)\right)^2
\sum_\alpha  c^\dagger_\alpha  c_\alpha 
\; .
\end{eqnarray}
Here $\delta\varphi(t)=(\phi_\omega/\phi_0)\sin(\omega t)$, 
and
$\hat P_x=-i d/dx +
(2\pi/L)(\phi/\phi_0)$ is the $x$-component of the
one particle momentum operator. 
As usual, the coordinate along the circumference
is called the $x$-direction, and we impose periodic
boundary conditions.
Using standard non-equilibrium Green's function methods,
the contribution to the
non-equilibrium current that is quadratic in the
external field is easily obtained\cite{Fricke95}:
\begin{eqnarray}
I^{(2)}(t) & = & \frac{(-e)(2\pi)^2}{(m L)^3} \int^\infty_\infty d\omega_1
d\omega_2 \delta\varphi_{\omega_1}\delta\varphi_{\omega_2} \nonumber
\\
&\times& K^{(2)} (\omega_1, \omega_2) e^{-i(\omega_1+\omega_2)t}
\label{eq:strom}
\; ,
\end{eqnarray}
where  $\varphi_{\omega}$ is the Fourier transform
of the time-dependent part of the
flux (\ref{eq:flux}) in units of the flux quantum
(i.e.
$\phi(t) - \phi = \phi_0 
\int d\omega' \delta\varphi_{\omega'} e^{-i\omega' t}$)
and
the response function $K^{(2)} ( \omega_1 , \omega_2 )$
is given by
\begin{eqnarray}
 K^{(2)}(\omega_1,\omega_2) &=& \sum_{\alpha\beta\gamma}
 \frac{P_{\alpha\beta\gamma}}{\varepsilon_\gamma-\varepsilon_\alpha
 +\omega_1 +\omega_2
 +i0}  \nonumber \\ 
 & & \hspace{-20mm} \times \biggl[\frac{f(\varepsilon_\gamma)-
 f(\varepsilon_\beta)}{\varepsilon_\gamma-
 \varepsilon_\beta +\omega_2 +i0} 
 -\frac{f(\varepsilon_\beta)-
 f(\varepsilon_\alpha)}{\varepsilon_\beta-\varepsilon_\alpha
  +\omega_1 +i0}\biggr] \; ,
 \label{eq:k2adiabatic}
 \end {eqnarray}
with 
 \begin{equation}
 \label{Pabc}
 P_{\alpha\beta\gamma}=\langle\alpha|\hat
 P_x|\beta\rangle\langle\beta|\hat P_x|\gamma\rangle\langle\gamma|\hat
 P_x|\alpha\rangle
 \; .
 \end{equation}
Here $f(\varepsilon_\alpha)= 
\langle  c^\dagger_\alpha c_\alpha\rangle $ is the
occupation number, which in a grand-canonical ensemble
is the Fermi function. 
Keeping in mind that the time-dependent part
of the flux (\ref{eq:flux}) corresponds to
 \begin{equation}
 \delta\varphi_{\omega'} =
 \frac{\phi_\omega}{2i\phi_0}[\delta(\omega'+\omega) -
 \delta(\omega'-\omega)]
 \; ,
 \end{equation}
it is clear that in this case
Eq.(\ref{eq:strom}) contains not only 
oscillating terms, but also a time-independent
contribution, 
 \begin{equation}
 I^{(2)}_0= A_{\omega}
  \bigl[ K^{(2)}(\omega,-\omega)
 + K^{(2)}(-\omega,\omega)\bigr]
 \; ,
 \label{eq:I2_0}
 \end{equation}
where
 \begin{equation}
 A_{\omega}
 =
 \frac{(-e)(2\pi\phi_\omega)^2}{4 (L
 m)^3 \phi_0^2} 
 \label{eq:Aomegadef}
 \; ,
 \end{equation}
and
 \begin{eqnarray}
 K^{(2)}(\omega,-\omega) &=& \sum_{\alpha\beta\gamma}
 \frac{P_{\alpha\beta\gamma}}{\varepsilon_\gamma-\varepsilon_\alpha
 +i0 }  \nonumber \\ 
 & & \hspace{-20mm} \times \biggl[\frac{f(\varepsilon_\gamma)-
 f(\varepsilon_\beta)}{\varepsilon_\gamma-\varepsilon_\beta
  -\omega +i0}  
 -\frac{f(\varepsilon_\beta)-f(\varepsilon_\alpha)}{
 \varepsilon_\beta-\varepsilon_\alpha
  +\omega +i0}\biggr]
  \; .
 \label{eq:k20}
 \end{eqnarray}
Defining retarded and advanced Green's functions,
 \begin{equation}
  G^{R}_\alpha(\varepsilon)=\frac{1}{
  \varepsilon - \varepsilon_\alpha + i0}
 \; \; , \; \; 
  G^{A}_\alpha(\varepsilon)=\frac{1}{
  \varepsilon - \varepsilon_\alpha - i0}
  \; ,
  \label{eq:green}
 \end{equation}
Eq.(\ref{eq:k20}) can also be written as
 \begin{eqnarray}
  K^{(2)}(\omega,-\omega)&=&-\frac{1}{2\pi
    i}\sum_{\alpha\beta\gamma} P_{\alpha\beta\gamma}
 \nonumber \\
 & & \hspace{-23mm} \times
 \Biggl\{ 
 \int_\infty^\infty
  d\varepsilon
  f(\varepsilon+\omega)  
 \bigl[ G^R_\alpha(\varepsilon+\omega)G^R_\beta(
 \varepsilon)G^R_\gamma(\varepsilon+\omega)
 \nonumber
 \\
 & & 
  \hspace{7mm} -
 G^A_\alpha(\varepsilon+\omega)G^A_\beta(
 \varepsilon)G^A_\gamma(\varepsilon+\omega) \bigr]
 \nonumber \\
 & & \hspace{-20mm} -
 \int_\infty^\infty
  d\varepsilon
  [f(\varepsilon+\omega)-f(\varepsilon)] \nonumber \\
 & \times & \bigl[ G^R_\alpha(\varepsilon+\omega)
 G^A_\beta(\varepsilon)G^A_\gamma(\varepsilon+\omega)
 \nonumber \\
 & & -
 G^R_\alpha(\varepsilon+\omega)G^R_\beta(
 \varepsilon)G^A_\gamma(\varepsilon+\omega) \bigr]\Biggr\} \; .
  \label{eq:k20green}
 \end{eqnarray}
The structure of the Green's functions agrees with the one given by
KY in Ref.\cite{Kravtsov97}.
Note, however, that these authors work in a different gauge:
they represent 
the electric field by a scalar potential, so that their
expressions contain only a single current vertex.
The introduction of Green's function is useful
for calculating disorder averages.
It is common wisdom that for the calculation of the
disorder average
of Eq.(\ref{eq:k20green})
the terms involving products of only retarded or only advanced
Green's functions can be neglected\cite{common}.
In this approximation a perturbative 
calculation of the disorder average 
of Eq.(\ref{eq:k20green}) has been given by
KY\cite{Kravtsov93}, with the result that the associated time-independent
part of the non-equilibrium current is proportional
to $\omega^{-2}$ for frequencies larger than the Thouless energy.
As explained in Sec.\ref{sec:intro}, for frequencies $\omega<E_c$ the
perturbative expansion is not controlled anymore since the energy
spectrum becomes discrete. In fact, it will turn out, that the physical
behavior is completely different in this regime. 

To demonstrate the break down of the diagrammatic perturbation theory
for systems with a discrete spectrum, we now show that {\it{an exact evaluation of the disorder
average of
Eq.(\ref{eq:k20green})}} should actually yield an {\it{infinite}} result.
Let us therefore go back
to the exact spectral representation (\ref{eq:k20})
of the response function.
Using the formal identity
 \begin{equation}
 \label{eq:i0id}
 \frac{1}{x+i0}=\wp\frac{1}{x}-i\pi\delta(x)
 \; ,
 \end{equation}
where $\wp$ denotes the Cauchy principal part,
we can rewrite Eq.(\ref{eq:k20}) as
 \begin{equation}
 \label{k20b}
 K^{(2)}(\omega,-\omega)=
 K^{(2)}_\wp(\omega,-\omega)+
 K^{(2)}_{\delta\delta}(\omega,-\omega)
 \; ,
 \end{equation}
with
 \begin{eqnarray}
 K^{(2)}_\wp(\omega,-\omega)&=& 
 2\sum_{\alpha\beta\gamma}
 \frac{ {\rm Re} P_{\alpha\beta\gamma}}{
 \varepsilon_\gamma-\varepsilon_\alpha } 
 \wp\frac{f(\varepsilon_\gamma)-f(
 \varepsilon_\beta)}{\varepsilon_\gamma-\varepsilon_\beta
 -\omega }
 \; ,
  \label{eq:K2P}
 \\
 K^{(2)}_{\delta\delta}(\omega,-\omega)&=&
 -2\pi^2\sum_{\alpha\beta\gamma} {\rm Re}
 P_{\alpha\beta\gamma}
 [f(\varepsilon_\gamma)-f(\varepsilon_\beta)]\nonumber\\
 & \times & \delta(\varepsilon_\gamma-\varepsilon_\alpha)
 \delta(\varepsilon_\gamma-\varepsilon_\beta-\omega)
 \label{eq:K2deltadelta}
 \; .
 \end{eqnarray}
The terms  with $\alpha = \gamma$
in Eqs. (\ref{eq:K2P}) and (\ref{eq:K2deltadelta})
yield the following contributions,
 \begin{eqnarray}
 K^{(2)}_{\wp,{\rm diag}}(\omega,-\omega) &=& 
 \wp
 \sum_{\alpha\beta}
 P_{\alpha\beta\alpha}\frac{\partial}{\partial\varepsilon_\alpha} 
 \frac{f(\varepsilon_\alpha)-
 f(\varepsilon_\beta)}{\varepsilon_\alpha-\varepsilon_\beta
 -\omega}
 \nonumber
 \\
 &  &  \hspace{-25mm} = \wp  \sum_{\alpha\beta}
 P_{\alpha\beta\alpha} \Biggl[
 \frac{ \frac{\partial}{\partial 
 \varepsilon_{\alpha} } f(\varepsilon_\alpha)
 }{\varepsilon_\alpha-\varepsilon_\beta
 -\omega}  
 - \frac{f(\varepsilon_\alpha)-
 f(\varepsilon_\beta)}{(\varepsilon_\alpha-\varepsilon_\beta
  -\omega)^2} \Biggr]
 \;  ,
 \label{eq:Kp_diag}
 \end{eqnarray}
 \begin{eqnarray}
 K^{(2)}_{\delta\delta, {\rm diag}}(\omega,-\omega)&=&
 -2\pi^2\delta(0)\sum_{\alpha\beta} {\rm Re}
 P_{\alpha\beta\alpha}[f(\varepsilon_\alpha)-
 f(\varepsilon_\beta)]\nonumber\\
 & & \times \delta(\varepsilon_\alpha-\varepsilon_\beta-\omega)  
 \; .
 \label{eq:k2deltadiag}
 \end{eqnarray}
The right-hand side of Eq.(\ref{eq:k2deltadiag}) is
proportional to the infinite
factor $\delta ( 0 )$. Hence,
the term
$K^{(2)}_{\delta\delta }(\omega,-\omega)$  
must also be infinite. Because the singular prefactor
$\delta ( 0)$ in Eq.(\ref{eq:k2deltadiag})
does not depend on the disorder,
this singularity survives disorder averaging\cite{dos}.
Keeping in mind that Eq.(\ref{eq:k20green})
is mathematically equivalent with Eq.(\ref{eq:k20}),
we conclude that
a correct evaluation of the disorder average 
$\overline{ K^{(2) } ( \omega , - \omega )}$
must yield an infinite result\cite{dos}.
Unfortunately, in an approximate evaluation of
Eq.(\ref{eq:k20green}) by means of the 
usual diagrammatic methods this $\delta$-function
singularity is
artificially smoothed out, and one obtains a finite 
result\cite{Kravtsov93}.

\section{Adiabatic switching on}
\label{sec:adon}

The infinite term (\ref{eq:k2deltadiag})
is clearly unphysical. This term 
is closely related to the
infinitesimal imaginary parts
$i0$ that have been added to the
real frequencies in the spectral representation
(\ref{eq:k20}) for the response function
$K^{(2)} ( \omega , - \omega)$. As emphasized by KY\cite{Kravtsov97},
the infinitesimal imaginary parts 
are a consequence of the fact that
the response function must be causal
when the  time-dependent part of the Hamiltonian
is adiabatically switched on.
Let us examine the
''adiabatic switching on'' of the time-dependent
perturbation more carefully. Following the
usual recipe\cite{Baym69},
we replace the Hamiltonian
$\hat H_0 + \hat V (t)$  by
$\hat H_0 + \hat V_\eta (t)$, where $\hat V_\eta (t)=\exp(\eta t) \hat
V(t)$. The limit $\eta \rightarrow 0$ is then taken at the end of the
calculation  of physical quantities.
For large enough times $t$ the physical result  
should be
independent of the switching on procedure. 
Indeed, in the appendix
we show by explicit calculation
that sudden switching on produces the same result
for the long-time response as adiabatic switching on.
However, in the latter case one still has to be careful
to take the limit $\eta
\rightarrow 0$ only after the physical quantity of interest
has been calculated. We now show that the 
singularity in Eq.(\ref{eq:k2deltadiag}) has been 
artificially generated by taking the limit $\eta \rightarrow 0$
at an intermediate step of the calculation.

By direct expansion of the time evolution operator
in the interaction representation
to second order in the time-dependent perturbation, we
obtain the current for adiabatic switching on with finite $\eta$
 \begin{eqnarray}
 I^{(2)}_\eta (t) & = & 
 \frac{(-e)(2\pi)^2}{(m L)^3} \int^\infty_\infty d\omega_1
 d\omega_2 \delta\varphi_{\omega_1}\delta\varphi_{\omega_2} \nonumber
 \\
 & \times  & K^{(2)}_{\eta t} (\omega_1, \omega_2) e^{-i(\omega_1+\omega_2)t}
 \; ,
 \label{eq:strometa}
 \end{eqnarray}
with
 \begin{eqnarray}
 K^{(2)}_{\eta t} (\omega_1,\omega_2) 
 & = & e^{2 \eta t} \sum_{\alpha\beta\gamma}
 \frac{P_{\alpha\beta\gamma}}{
 \varepsilon_\gamma-\varepsilon_\alpha
 +\omega_1 +\omega_2
 +2i\eta}  \nonumber \\ 
 & & \hspace{-20mm} \times 
 \biggl[\frac{f(\varepsilon_\gamma)-
 f(\varepsilon_\beta)}{\varepsilon_\gamma-\varepsilon_\beta
 +\omega_2 +i\eta}
 - \frac{f(\varepsilon_\beta)-
 f(\varepsilon_\alpha)}{\varepsilon_\beta-\varepsilon_\alpha
 +\omega_1 +i\eta}\biggr] \; .
 \label{eq:k2eta}
\end {eqnarray}
Comparing Eq.(\ref{eq:k2eta}) with
Eq.(\ref{eq:k2adiabatic}), 
we see that the former is multiplied by an extra factor of
$e^{ 2 \eta t }$.  
If we directly take the limit $\eta \rightarrow 0$,
this factor is replaced by unity.
This is the limiting procedure adopted in 
the usual Green's
function approach,  where
one takes first the limit
$\eta\rightarrow 0$ in Eq.(\ref{eq:k2eta}) and then inserts the
resulting expression into Eq.(\ref{eq:strometa}). 
In this case we recover
Eqs.(\ref{eq:I2_0}) and (\ref{eq:k20}), which lead to
the divergence in Eq.(\ref{eq:k2deltadiag}). 
We now show that this unphysical divergence
does not appear if the limit $\eta \rightarrow 0$
is taken {\it{after the physical current has been calculated}}.
Substituting Eq.(\ref{eq:k2eta}) 
into Eq.(\ref{eq:strometa}) we obtain 
 \begin{eqnarray}
 I^{(2)}_\eta (t) & = & A_{\omega}
 \bigl[
  K^{(2)}_{\eta t}(\omega,-\omega) +
  K^{(2)}_{\eta t}(- \omega, \omega)
 \nonumber
 \\
 &  & 
 + K^{(2)}_{\eta t}(\omega,\omega)e^{-2i \omega t}
 + K^{(2)}_{\eta t}(- \omega, - \omega)e^{2i \omega t}
 \bigr]
 \; .
 \label{eq:I2_eta2}
 \end{eqnarray} 
In analogy with Eq.(\ref{k20b}),  we express
$K^{(2)}_{\eta t}(\omega,-\omega)$ in terms of products of real 
and imaginary parts
 \begin{equation}
 \label{k2_etab}
 K^{(2)}_{\eta t} (\omega,-\omega)=K^{(2)}_{\eta
 t, \wp}(\omega,-\omega)+K^{(2)}_{\eta t, \delta \delta}(\omega,-\omega)
 \; ,
 \end{equation}
with 
 \begin{eqnarray}
 K^{(2)}_{\eta t, \wp} (\omega,-\omega) & = & 
 2 
 e^{2\eta t} 
 \sum_{\alpha\beta\gamma}
 {\rm Re}
 P_{\alpha\beta\gamma}[f(\varepsilon_\gamma)-
 f(\varepsilon_\beta)] \nonumber \\
 & & \hspace{-20mm} \times 
 \biggl[\frac{\varepsilon_\gamma-\varepsilon_\alpha}{
 (\varepsilon_\gamma-\varepsilon_\alpha)^2
 +(2\eta)^2} \frac{\varepsilon_\gamma-\varepsilon_\beta
 -\omega}{(\varepsilon_\gamma-\varepsilon_\beta -\omega)^2 +
 \eta^2}\biggr]  \; ,  
 \label{eq:k2_eta_rr}
 \end{eqnarray}
 \begin{eqnarray}
 K^{(2)}_{\eta t, \delta \delta }(\omega,-\omega) & = &
 -2 
 e^{2\eta t} 
 \sum_{\alpha\beta\gamma}
 {\rm Re}
 P_{\alpha\beta\gamma}[f(\varepsilon_\gamma)-f(\varepsilon_\beta)]
 \nonumber
 \\
 & & \hspace{-20mm} \times
 \biggl[ \frac{2\eta}{(\varepsilon_\gamma-\varepsilon_\alpha)^2
  +(2\eta)^2}  
 \frac{\eta}{(\varepsilon_\gamma-\varepsilon_\beta -
 \omega)^2 + \eta^2}\biggr]
 \; .
 \label{eq:k2_eta_ii}
 \end {eqnarray} 
From Eq.(\ref{eq:k2_eta_rr}) it is now obvious that
$K^{(2)}_{\eta t, {\wp} }$ does not have any contributions
from the terms 
$\alpha=\gamma$. The finite contribution 
in Eq.(\ref{eq:Kp_diag})
is thus an artifact of taking the limit $\eta\rightarrow
0$ before calculating any physical quantities. 
Let us now focus on the term
(\ref{eq:k2_eta_ii}).
If we directly take the limit
$\eta \rightarrow 0$ 
using
 \begin{equation}
 \lim_{\eta \rightarrow 0} \frac{ \eta}{\epsilon^2 + \eta^2}
 = \pi \delta ( \epsilon )
 \; ,
 \end{equation}
we recover the infinite result (\ref{eq:k2deltadiag}).
However, the structure of the $\eta$-dependent part
of Eq.(\ref{eq:k2_eta_ii}) is familiar 
from the derivation of Fermi's golden rule of elementary
quantum mechanics.
As discussed for example in the classic textbook by Baym\cite{Baym69},
terms with this structure should be
interpreted as a {\it{rate}},  
i.e. as a contribution to the current that grows linearly in time.
It is therefore clear that after taking
the derivative of Eq.(\ref{eq:k2_eta_ii}) with respect to
$t$ we obtain a finite result if we then let
$\eta \rightarrow 0$.
A simple calculation yields
 \begin{eqnarray}
 \lim_{\eta\rightarrow 0} \frac{d}{dt} K^{(2)}_{\eta
 t, \delta \delta }(\omega,-\omega) & &
 \nonumber
 \\
 &  & \hspace{-30mm} =
 -2 \lim_{\eta\rightarrow 0}
 \sum_{\alpha\beta}P_{\alpha\beta\alpha}
 \frac{ [f(\varepsilon_\alpha)-
 f(\varepsilon_\beta)] \eta }{(\varepsilon_\alpha-
 \varepsilon_\beta-\omega)^2+\eta^2}
 \nonumber\\
 &  & \hspace{-30mm} = 
 -2\pi 
 \sum_{\alpha\beta}P_{\alpha\beta\alpha}
 [f(\varepsilon_\alpha)-f(\varepsilon_\beta)]
 \delta(\varepsilon_\alpha-\varepsilon_\beta-\omega)   
 \; .
 \label{eq:k_eta0_ii}
 \end{eqnarray}
Because this expression contains only a single $\delta$-function,
after averaging over disorder it becomes a smooth function
of $\omega$.
We conclude that to quadratic order in the field
the non-equilibrium current
induced by the time-dependent
flux (\ref{eq:flux}) has the following three contributions,
 \begin{equation}
 I^{(2)} ( t ) \equiv \lim_{\eta \rightarrow 0}
 I^{(2)}_{\eta} ( t ) =
 I^{(2)}_{\rm th}  +  t \frac{d I^{(2)}_{{\rm kin}}}{dt}  + I^{(2)}_{\rm osc} (t )
 \; ,
 \label{eq:Ifinal}
 \end{equation}
where the time-independent  part is given by
 \begin{eqnarray}
 I^{(2)}_{\rm th}  
 & =  & A_{\omega}
 \lim_{\eta \rightarrow 0}
 \bigl[
  K^{(2)}_{\eta t, \wp }(\omega,-\omega) +
  K^{(2)}_{\eta t, \wp }(- \omega, \omega)
  \bigr]
 \nonumber
 \\
 & =  & 2 A_{\omega}
 \sum_{\alpha\beta\gamma, \alpha \neq \gamma }
 \frac{ {\rm Re} P_{\alpha\beta\gamma}}{
 \varepsilon_\gamma-\varepsilon_\alpha } 
 \nonumber
 \\
 & \times  & 
 \wp \left[
 \frac{f(\varepsilon_\gamma)-f(
 \varepsilon_\beta)}{\varepsilon_\gamma-\varepsilon_\beta
 -\omega }
 + ( \omega \rightarrow - \omega ) \right]
 \; .
 \label{eq:Ith}
 \end{eqnarray}
The coefficient of the term linear in time is
 \begin{eqnarray}
 \frac{ d I^{(2)}_{\rm kin}  }{dt}
 & =  & A_{\omega}
 \lim_{\eta \rightarrow 0}
 \bigl[
   \frac{d}{dt} K^{(2)}_{\eta t, \delta \delta  }(\omega,-\omega) +
  \frac{d}{dt} K^{(2)}_{\eta t, \delta \delta  }(- \omega, \omega)
  \bigr]
 \nonumber
 \\
 & = & - 2 \pi A_{\omega}
 \sum_{\alpha\beta}P_{\alpha\beta\alpha}
 [f(\varepsilon_\alpha)-f(\varepsilon_\beta)]
 \nonumber
 \\
 & & \times
 \left[ \delta(\varepsilon_\alpha-\varepsilon_\beta-\omega)   
 + ( \omega \rightarrow - \omega ) \right]
 \; ,
 \label{eq:Ikin}
 \end{eqnarray}
and the oscillating part is 
 \begin{equation}
 I^{(2)}_{\rm osc} (t)  =  A_{\omega}
 \lim_{\eta \rightarrow 0} \bigl[
 K^{(2)}_{\eta t}(\omega,\omega) e^{- 2 i \omega t} 
 +
 K^{(2)}_{\eta t}(- \omega, - \omega) e^{2 i \omega t}
 \bigr]
 \; . 
 \end{equation}
Thus,  
a time-dependent electric field with
frequency $\omega $ induces in quadratic order
three fundamentally different currents.
(a) A time-independent contribution $I^{(2)}_{\rm th}$;
as shown in the next section, this contribution
can be derived from a thermodynamic calculation.
(b) A contribution $t d I^{(2)}_{\rm kin}/ dt$
which increases linearly in time;
this term can be understood 
in terms of the usual golden rule
of time-dependent perturbation theory.
(c) Finally, there is also a time-dependent contribution $I^{(2)}_{\rm osc}$ 
oscillating with frequency $2 \omega$. 
When this term is averaged over a time-interval
larger than $\omega^{-1}$, its contribution
to the current is negligibly small.

From the above analysis it is clear that the contribution 
that is proportional to $t$ cannot be calculated 
within the usual Green's function machinery, because
in this approach the limit $\eta \rightarrow 0$ is
taken at an intermediate step of the calculation, causing
an unphysical divergence.
To further support the correctness  
of the limiting procedure adopted here 
we show in the appendix that
Eqs.(\ref{eq:Ifinal}--\ref{eq:Ikin}) 
can also be re-derived if the
perturbation is {\it{suddenly}} (instead of
adiabatically) switched on.

\section{The thermodynamic origin of  the time-independent
part of the current}

The time-independent part $I^{(2)}_{\rm th}$ of the
non-equilibrium current in Eq.(\ref{eq:Ifinal}) has
been discussed by us in Ref.\cite{Kopietz97}.
This contribution 
can be obtained from a thermodynamic calculation.
In Ref.\cite{Kopietz97}
we have assumed (without further justification)
the existence of such a relation.
Let us now put this assumption
on a more solid basis.
Within the Matsubara (imaginary time) formalism one 
can directly calculate the imaginary frequency version
of the response function $K^{(2) } ( \omega_1 , \omega_2 )$
given in Eq.(\ref{eq:k2adiabatic}), i.e.
 \begin{eqnarray}
 K^{(2)}(i\omega_1,i\omega_2) &=& \sum_{\alpha\beta\gamma}
 \frac{P_{\alpha\beta\gamma}}{\varepsilon_\gamma-\varepsilon_\alpha
 +i\omega_1 +i\omega_2
 }  \nonumber \\ 
 &\times & \biggl[ \frac{f(\varepsilon_\gamma)-
 f(\varepsilon_\beta)}{\varepsilon_\gamma-\varepsilon_\beta +i\omega_2} 
 -\frac{f(\varepsilon_\beta)-
 f(\varepsilon_\alpha)}{\varepsilon_\beta-\varepsilon_\alpha
 +i\omega_1}\biggr] \; .   
 \label{k2iomega}
 \end{eqnarray}
As pointed out by KY\cite{Kravtsov97},
in order to obtain the causal response function, one should
first continue both frequencies to the real axis
with positive imaginary part
($ i \omega_1 \rightarrow \omega_1 + i 0$,
$ i \omega_2 \rightarrow \omega_2 + i 0$), and 
then set $\omega_1 = - \omega_2$ to obtain the
constant part of the physical current.
On the other hand, if one performs these steps
in opposite order
(i.e. first sets $i \omega_1 = - i \omega_2$ and then 
continues $i \omega_1 \rightarrow \omega + i 0$) one obtains
for the current response function
 \begin{eqnarray}
 K^{(2)}_{{\rm th}}(\omega,-\omega) & =& {\rm Re} 
 \sum_{\alpha\beta\gamma}
 \frac{P_{\alpha\beta\gamma}}{\varepsilon_\gamma-\varepsilon_\alpha
  }  \nonumber \\ 
 & & \hspace{-15mm} \times  \biggl[\frac{f(\varepsilon_\gamma)-
 f(\varepsilon_\beta)}{\varepsilon_\gamma-\varepsilon_\beta
 -\omega -i0} 
 -\frac{f(\varepsilon_\beta)-f(\varepsilon_\alpha)}{
 \varepsilon_\beta-\varepsilon_\alpha
  +\omega +i0}\biggr]  
  \; .
 \label{eq:k2iomega0}
 \end{eqnarray}
Comparing this expression with Eqs.(\ref{k20b}--\ref{eq:K2deltadelta}),
it is easy to see that
 \begin{equation}
 K^{(2)}_{\rm th} ( \omega , - \omega ) =
 K^{(2)}_{\wp} ( \omega , - \omega )
 \; .
 \end{equation}
Hence, the time-independent part $I^{(2)}_{\rm th}$ of the current 
can indeed be
obtained  from a thermodynamic calculation\cite{Kopietz97}.
Note, however, that our analysis of Sec.\ref{sec:adon} 
(see also the appendix) implies that
the terms with $\alpha = \gamma$ in Eq.(\ref{eq:k2iomega0})
should be omitted from the sum, i.e. 
the physical current is given by
 \begin{equation}
 I_{\rm th}^{(2)} = A_{\omega}
 \left[ \tilde{K}_{\rm th}^{(2)} ( \omega , - \omega )
 + \tilde{K}_{\rm th}^{(2)} ( - \omega ,  \omega )
 \right]
 \; .
 \end{equation}
 where
 \begin{eqnarray}
 \tilde{K}^{(2)}_{{\rm th}}(\omega,-\omega) & =& 
 {K}^{(2)}_{ \wp }(\omega,-\omega) - 
 {K}^{(2)}_{\wp , {\rm diag} }(\omega,-\omega)  
 \nonumber
 \\
 &   & \hspace{-10mm} =
 \sum_{\alpha\beta\gamma , \alpha \neq \gamma}
 \frac{P_{\alpha\beta\gamma}}{\varepsilon_\gamma-\varepsilon_\alpha
  }  \nonumber \\ 
 &  & \hspace{-10mm} \times  \wp \biggl[\frac{f(\varepsilon_\gamma)-
 f(\varepsilon_\beta)}{\varepsilon_\gamma-\varepsilon_\beta
 -\omega } 
 -\frac{f(\varepsilon_\beta)-f(\varepsilon_\alpha)}{
 \varepsilon_\beta-\varepsilon_\alpha
  +\omega }\biggr]  
  \; ,
 \label{eq:k2iomegath}
 \end{eqnarray}
see Eq.(\ref{eq:Kp_diag}).
The direct diagrammatic calculation 
of the disorder average of
${I^{(2)}_{\rm th}}$ is difficult, because
the restriction $\alpha \neq \gamma$ 
in Eq.(\ref{eq:k2iomegath}) is not so easy
to implement.
In Ref.\cite{Kopietz97} the
following limiting procedure was adopted:
Instead of directly calculating
 $ \overline{\tilde{K}^{(2)}_{{\rm th}}(\omega,-\omega)}$,
consider  the generalization
of the imaginary frequency
response function (\ref{k2iomega}) for
electric fields with finite wave-vector ${\bf{q}}$,
which we denote by
$\overline{K^{(2)} ( i \omega , - i \omega , {\bf{q}} )}$.
The limit ${\bf{q}} \rightarrow 0$ is taken after the
disorder averaged current has been calculated.
As shown in Ref.\cite{Kopietz97}, in the diffusive regime
the function
$\overline{K^{(2)} ( i \omega , - i \omega , {\bf{q}} )}$
is a smooth function of ${\bf{q}}$, so that
the limit ${\bf{q}} \rightarrow 0$ is well defined.
The so-defined averaged response function
vanishes for  frequencies exceeding the Thouless energy
as $\exp ( - \sqrt{ | \omega | / 2 E_c } )$\cite{Kopietz97}.
On the other hand, 
perturbative averaging of the
contribution from the (unwanted) diagonal
term (\ref{eq:Kp_diag}) shows that
this term vanishes as $ \omega^{-2}$ for large frequencies.
This indicates that the above limiting procedure indeed
eliminates the contribution of the unphysical
diagonal term
(\ref{eq:Kp_diag})
to the time-independent part of the non-equilibrium current.

\section{Conclusion}
In this work we have shown that
a time-dependent 
flux oscillating with frequency $\omega$ that pierces the center 
of a
mesoscopic metal ring generates to quadratic order 
three fundamentally different contributions to the current:
a constant
non-equilibrium current $I^{(2)}_{\rm th}$, 
a current $t d I^{(2)}_{\rm kin} / dt$ that grows linearly in time, and
a current oscillating with frequency $2 \omega$.
As shown in Ref.\cite{Kopietz97},
the 
disorder average of the constant term 
$\overline{I^{(2)}_{\rm th}}$ 
vanishes 
for frequencies exceeding the Thouless energy 
as $\exp ( - \sqrt{ | \omega | / 2 E_c } )$.
The calculation of the disorder average of 
the contribution $t d I^{(2)}_{\rm kin} / dt$ 
remains an open problem. A direct perturbative calculation
by means of the impurity diagram technique is not straightforward, 
because Eq.(\ref{eq:Ikin}) involves three matrix elements but only
one energy denominator. Therefore this expression cannot be
simply written in terms of Green's functions.

The main result of this work is the prediction
of a current
$t d I^{(2)}_{\rm kin} / dt$ increasing linearly
with time. From the well-known derivation of Fermi's golden rule\cite{Baym69}
it is clear that this result is only valid in an intermediate
time interval. In particular, the calculation
of the long-time behavior of the non-equilibrium current
requires non-perturbative methods.

One should keep in mind that our calculation has been
performed for non-interacting electrons in a random potential,
so that our results are valid as long as 
the spectrum of the system is discrete. We have argued in Sec.\ref{sec:intro}
that at low enough temperatures this should be the case for small
external frequencies, $ | \omega | <E_c$. 
On the other hand, 
for frequencies exceeding $E_c$ the
spectrum is effectively continuous.
In this regime
the conventional Green's function methods can be used
to calculate the direct current, 
so that the results of
KY\cite{Kravtsov93} should be valid.

Let us also point out that the linear time-dependence of the current
is a consequence of the discrete spectrum, and is not related
to the adiabatic switching on procedure in Eq.(\ref{eq:k2eta}).
In the appendix we show that sudden switching on
yields the same linear time-dependence of the current.
It seems reasonable to expect that for sufficiently short times the
constant part $I^{(2)}_{\rm th}$ of the current is
dominant\cite{Kopietz97}. 
We would like to encourage experimentalists to
measure
the non-equilibrium response of mesoscopic metal rings
to a time-dependent flux in the frequency
range $10^8 {\rm Hz} \leq \omega \leq 10^{13} {\rm Hz}$.

\section*{Acknowledgement}

This work was supported by the
Deutsche Forschungsgemeinschaft (SFB 345).
We thank V. E. Kravtsov for his comments.

\begin{appendix}
\section*{Sudden switching on}
\label{appendixA}
To confirm that the ''switching on procedure'' outlined in 
Sec.\ref{sec:adon} yields the correct physical results, let us
consider a
harmonic perturbation that is
suddenly turned on at time $t=0$, 
 \begin{equation}
 \phi ( t )  =
 \phi + \phi_\omega  \Theta(t)\sin ( \omega t )
 \; ,
 \label{eq:flux2}
 \end{equation}
where $\Theta(t)$ is the step function. 
To second order in $\phi_{\omega}$ the induced current is
 \begin{eqnarray}
 \label{eq:stromein}
 I^{(2)}(t)&=&
 2 A_{\omega} {\rm Re}
 \sum_{\alpha\beta\gamma}
 P_{\alpha\beta\gamma}[f(\varepsilon_\beta)-f(\varepsilon_\alpha)]
 \nonumber \\
 &  & \hspace{-15mm} \times
 \Biggl[ \frac{ e^{2i\omega
 t}-e^{i(\varepsilon_\gamma-
 \varepsilon_\alpha)t}}{(\varepsilon_\alpha-
 \varepsilon_\gamma+2\omega)(\varepsilon_\alpha-
 \varepsilon_\beta +\omega )} 
 \nonumber
 \\
 & & \hspace{-10mm}
 -
 \frac{1-e^{i(\varepsilon_\gamma-\varepsilon_\alpha)t}}{
 (\varepsilon_\alpha-
 \varepsilon_\gamma)(\varepsilon_\alpha-\varepsilon_\beta-\omega)}
 \nonumber \\
 & & \hspace{-10mm} +
 \frac{2 \omega }{ ( \varepsilon_\alpha-\varepsilon_\beta )^2
- \omega^2 }
 \Biggl[ \frac{e^{i(\varepsilon_\beta-\varepsilon_\alpha+
\omega)t}-e^{i(\varepsilon_\gamma-\varepsilon_\alpha)t}}{
\varepsilon_\beta-\varepsilon_\gamma
+\omega}  \Biggr]
\nonumber
\\
& & \hspace{-10mm}
+ (\omega\rightarrow -\omega) \Biggr]
\; .
\end{eqnarray}
The diagonal term $\alpha=\gamma$ is
 \begin{eqnarray}
 \label{eq:idiag}
 I^{(2)}_{\rm diag}(t)&=&  4 A_{\omega}
 \sum_{\alpha\beta}
 P_{\alpha\beta\alpha}[f(\varepsilon_\alpha)-
 f(\varepsilon_\beta)] \nonumber \\
 & \times  & \Biggl[  \frac{ \sin^2(\omega
 t)}{  ( \varepsilon_\alpha-\varepsilon_\beta )^2 
  - \omega^2 }
  \nonumber \\
 & &  -  \frac{
 \sin^2\bigl(\frac{\varepsilon_\beta-\varepsilon_\alpha
  +\omega}{2}t\bigr) 
 + \sin^2\bigl(\frac{\varepsilon_\beta-\varepsilon_\alpha
  -\omega}{2}t\bigr)}{
 ( \varepsilon_\alpha-\varepsilon_\beta)^2-
 \omega^2 }
 \nonumber \\
 & & + 
 \Bigl[ \frac{\sin \bigl(\frac{\varepsilon_\beta-\varepsilon_\alpha
 +\omega}{2}t\bigr)}{\varepsilon_\beta-\varepsilon_\alpha
 +\omega} \Bigr]^2 +
 \Bigl[ \frac{\sin \bigl(\frac{
 \varepsilon_\beta-\varepsilon_\alpha
 -\omega}{2}t\bigr)}{\varepsilon_\beta-\varepsilon_\alpha
 -\omega} \Bigr]^2 \Biggr] 
 \; .
 \end{eqnarray}
The terms in the last line can be interpreted in the same way as is
done in Fermi's golden rule\cite{Baym69} by using the identity 
\begin{equation}
\label{eq:sindelta}
\left[\frac{\sin\left(\frac{\Delta\varepsilon}{2}
t\right)}{\Delta\varepsilon}
\right]^2 \rightarrow \frac{ \pi}{2} t 
\delta(\Delta\varepsilon) \; \textrm{for} 
\;\; t\rightarrow\infty
\; .
\end{equation}
It is now easy to see that for large times $I^{(2)}_{\rm diag}(t)$ 
yields exactly the
same linear in time contribution as given in
Eq.(\ref{eq:Ikin}). 
The terms with
no explicit time dependence in Eq.(\ref{eq:stromein}) can 
be identified with $I_{\rm th}^{(2)}$ in Eq.(\ref{eq:Ith}).

\end{appendix}

%

%
%

\begin{thebibliography}{99}
\bibitem{common}
Note, however, that in a canonical ensemble the chemical
potential fluctuates, so that disorder averages
involving only retarded or only advanced Green's functions
can also generate large contributions 
to response functions, see Ref.\cite{Fricke95}.
\bibitem{dos}
The same type of singularity occurs if one 
calculates
the disorder average
of the square of the density of states,
$\overline{ \rho^2 ( \epsilon ) } =
\overline{ 
\sum_{\alpha \beta } 
\delta ( \varepsilon - \varepsilon_{\alpha} )
\delta ( \varepsilon - \varepsilon_{\beta} )}$.
The contribution from the terms with $\alpha = \beta$
is proportional to $\delta ( 0 ) =  \infty$.
Yet, perturbative averaging with the help of the
impurity diagram technique produces a finite result.
\end{thebibliography}
\end{document}